\documentclass[pra,aps,twocolumn ,longbibliography]{revtex4-1}
\usepackage[colorlinks=true, citecolor=blue, urlcolor=orange ]{hyperref}
\usepackage{graphicx}% Include figure files
\usepackage{bm}
\usepackage{amsmath,amsfonts}
\usepackage{amsthm}
\usepackage{hyperref}
\usepackage{xcolor}

\usepackage{braket}
\theoremstyle{plain}
\usepackage{color}
\usepackage{amssymb}
\usepackage{amsthm}
\usepackage{amsfonts}
\usepackage{float}
\usepackage{tabularx}
\usepackage{graphicx}
\usepackage[export]{adjustbox}

\usepackage{mathtools}
\usepackage{esvect}
\usepackage{wrapfig}
\usepackage{amsthm}
\usepackage{verbatim}
\usepackage{bbm}
\usepackage[normalem]{ulem}

\usepackage{enumitem}
\usepackage{fmtcount}
\usepackage{booktabs}
\usepackage{csquotes}
\usepackage{epsfig}

\usepackage{tabularx}
\usepackage{graphicx}
\usepackage{amsmath}
\usepackage{braket}
\usepackage{latexsym}
\usepackage{bm}
\usepackage{graphics,epstopdf}
\usepackage{enumitem}
\usepackage{fmtcount}
\usepackage{booktabs}
\usepackage{csquotes}
\usepackage{epsfig}

\theoremstyle{plain}

\def\bea{\begin{eqnarray}}
\def\eea{\end{eqnarray}}
\def\ba{\begin{array}}
\def\ea{\end{array}}

\def\beq{\begin{equation}}
\def\eeq{\end{equation}}

\usepackage[normalem]{ulem}
\usepackage{float}
\usepackage{graphicx}  
\usepackage{dcolumn}          
\usepackage{amssymb}
\usepackage{appendix}
\usepackage{physics}   
\usepackage{mathtools}
\usepackage{esvect}
\usepackage{wrapfig}
\usepackage{amsthm}
\usepackage{verbatim}
\usepackage{bbm}

\usepackage[mathscr]{euscript}
\def\Tr{\operatorname{Tr}}

\def\({\left(}
\def\){\right)}
\def\[{\left[}
\def\]{\right]}

\newtheorem{theorem}{Theorem}

\begin{document}
\raggedbottom
\title{Precision Enhancement in Transient Quantum Thermometry:\\Cold-Probe Bias and Its Removal}

\title{Cold-Probe Bias in Transient Quantum Thermometry and Its Fate Under Environmental Memory}
%and its Breakdown under Environmental Memory}

\author{Debarupa Saha}
\author{Ujjwal Sen}
\affiliation{Harish-Chandra Research Institute, Chhatnag Road, Jhunsi, Prayagraj  211 019, India\\
Homi Bhabha National Institute, Training School Complex, Anushakti Nagar, Mumbai 400 094, India}

\begin{abstract}
We unveil a fundamental temperature bias in transient quantum thermometry under Markovian dynamics. For qubit probes evolving in a thermal Markovian environment, we prove that transient precision beyond the steady-state benchmark can be achieved if and only if the probe is initially colder than the bath temperature to be estimated. Cold probes are therefore both necessary and sufficient for enhanced transient precision in the Markovian regime. We then investigate the fate of this bias in the presence of environmental memory. In particular, in a non-Markovian scenario generated by an auxiliary-mediated system-bath coupling, we find that the cold-probe requirement for enhanced transient precision persists, indicating that the temperature bias survives certain forms of memory effects. In contrast, for a non-Markovian collisional model with perfect swap interactions between bath ancillas, transient enhancement is entirely absent regardless of the probe’s initial temperature. This indicates that strong non-Markovianity can lead to the complete disappearance of the enhancement effect, placing hot and cold probes on equal footing, with neither capable of achieving enhanced precision in this regime.

%We then investigate the robustness of this thermodynamic bias under non-Markovian dynamics. For structured non-Markovian scenarios in which memory arises through an auxiliary-mediated coupling, the necessity of cold probes remains robust within the explored parameter regimes, indicating that the cold-probe bias can persist even in non-Markovian settings. To assess whether this feature is universal, we consider a non-Markovian collisional model with perfect swap interactions. In this setting, the temperature bias disappears entirely: neither initially colder nor hotter probes provide any transient advantage over the thermal steady-state precision. Enhanced transient precision itself ceases to occur, demonstrating that the thermodynamic bias identified in the Markovian regime is not universal and can be eliminated by certain forms of environmental memory.
\end{abstract}
\maketitle
\section{Introduction}
Thermometry concerns the estimation of the temperature of a thermal bath. In classical thermometry, this is achieved using macroscopic devices, such as conventional thermometers. In contrast, quantum thermometry~\cite{QT1,Therm2013,QT2015,Therm2015, Therm20152, Therm20153,QT2016,Therm2017,Therm2018,Therm20182,QT2,QT3,Therm2019,QT4,QT5,QT7,Therm2021, QTherm2022,QT6,QT8,QT9,QT10,QT11,therm2025} employs quantum systems, referred to as probes, to estimate temperature.

Traditionally, temperature estimation was performed by measurements on the canonical equilibrium (thermal) state of the probe. This approach maintains an analogy with classical thermometry, where the temperature can be inferred from the equilibrium properties of the thermometer, rendering the estimation independent of its initial configuration. However, in classical thermometry, the temperature can be determined deterministically, thereby eliminating the need for a precision based analysis. In contrast, quantum thermometry exhibits intrinsically probabilistic behavior, and temperature estimation is characterized by a finite precision. Notably, recent studies have shown that non-equilibrium/non-canonical probe states can significantly enhance estimation precision~\cite{non2020,non2022,non20222,N1,non2023,N2,cofi,N3,N4,Can,Nonmark2024}, a feature that has also been experimentally demonstrated~\cite{Ex1,Ex2,Ex3,Ex4}.

As a consequence, the attainable precision depends sensitively on the choice of the initial probe state, since different initial preparations of the probe lead to different non-equilibrium evolutions, resulting in distinct estimation precision at transient times (before reaching equilibrium). Understanding which initial states lead to enhanced precision is therefore of both fundamental and practical importance in quantum thermometry.

In this work, we take a step in this direction by adopting a metrological approach~\cite{Metrev1,Metrev2,Metrev3,Metrev4,Metrev5,Metrev6,Metrev7,Metrev8} to quantum thermometry, wherein the bath temperature is encoded in the quantum probe through its interaction with the thermal bath. The temperature is then inferred via suitable measurements on the probe, and the ultimate precision is quantified using the quantum Fisher information (QFI)~\cite{SLD1,Braunstein1}. 

We consider a qubit probe, as qubits are the simplest nontrivial quantum systems and have attracted significant attention in quantum thermometry~\cite{Qu1,Qu2,Qu3,Qu4,Qubit2018,Qu5,cofi,Qu6,Qu7,Qu8,Qu9}, with their effectiveness demonstrated both theoretically and experimentally across various platforms~\cite{ExQu0,ExQu1,ExQu2,ExQu3,ExQu4}.

We first study the estimation of the bath inverse temperature in a setting where the temperature is encoded in the probe through its interaction with a bosonic bath, and the dynamics is Markovian~\cite{Ma1,Ma2,Mark1,gksl1,gksl2,RH,gksl3}. In this regime, we show that enhanced precision, quantified by the QFI, at transient times is achieved if and only if the probe is initially colder than the thermal state corresponding to the bath temperature. Here, colder and hotter probes refer to initial probe temperatures that are, respectively, lower or higher than that of the corresponding equilibrium or thermal state. Remarkably, this temperature bias persists for arbitrary choices of the probe Hamiltonian energy gap.

We next analyze how this temperature bias is affected in the presence of environmental memory, leading to non-Markovian dynamics~\cite{NMark1,NMark2,NMark3,NmArk4,NMArk5,NMark6}. We consider two distinct scenarios: (i) non-Markovianity induced via auxiliary-mediated interactions, and (ii) a quantum collisional model~\cite{CM1,CM2,CM3,CM4,CM5,CM6,CM7,CM8} under perfect swapping conditions, corresponding to a strongly non-Markovian regime. 

In the first case, we find that enhancement of precision over the thermal-state benchmark at transient times still requires the probe to be initially colder than the bath. This demonstrates that the bias toward colder probes persists under this setting of non-Markovian dynamics. In contrast, for the collisional model with strong memory effects, the enhancement of precision at transient times is entirely suppressed, irrespective of the initial state of the probe. As a result, both initially hotter and colder probes perform equivalently, indicating that the temperature bias disappears in this regime.

Thus, our results provide a clear and practical guideline for quantum thermometry by identifying when transient precision enhancement can be achieved and how it depends on both the initial probe state and the nature of the underlying dynamics. In particular, the condition derived in the Markovian regime offers a simple operational principle for designing efficient quantum thermometers, guaranteeing an advantage in precision over that obtained using the thermal state. Furthermore, the demonstrated sensitivity of this advantage to non-Markovian effects highlights the crucial role of environmental memory in quantum thermometry.

%These findings also reveal a clear operational distinction between Markovian and non-Markovian quantum thermometry. Under Markovian dynamics, transient precision enhancement occurs if and only if the probe is initially colder than the thermal state corresponding to the bath temperature, thereby establishing a necessary and sufficient condition. In contrast, strong non-Markovian effects can significantly modify or even suppress this advantage.

%These insights not only deepen the fundamental understanding of temperature estimation at the quantum level but also have direct implications for the development of quantum technologies that require fast and precise thermometry, such as nanoscale thermal sensing and quantum devices operating far from equilibrium.
%In contrast, non Markovian dynamics relax this constraint, allowing probes with different initial temperatures to achieve the same transient enhancement.

The rest of the paper is organized as follows. In Sec.~\ref{prem}, we discuss the preliminaries of quantum thermometry as well as Markovian and non Markovian dynamics. In Sec.~\ref{mar}, we analyze the Markovian case and demonstrate the existence of a bias favoring colder thermometers. In Sec.~\ref{non-mar}, we consider the case of auxiliary mediated non-Markovian evolution and show the persistence of the bias. In Sec.~\ref{noncol}, we analyze the non-Markovian quantum collisional model and show how the phenomenon of enhanced precision at transient times ceases to exist in this scenario. Finally, we conclude in Sec.~\ref{con}.
\section{Preliminaries}
\label{prem}
\subsection{Quantum Thermometry }
Quantum thermometry~\cite{QT1,Therm2013,QT2015,Therm2015, Therm20152, Therm20153,QT2016,Therm2017,Therm2018,Therm20182,QT2,QT3,Therm2019,QT4,QT5,QT7,Therm2021, QTherm2022,QT6,QT8,QT9,QT10,QT11,therm2025} addresses the problem of estimating the temperature
$T$, or equivalently the inverse temperature $\beta=1/T$, of a thermal bath $B$.
The basic setup consists of a probe system $S$, which may be a $d$-level quantum
system, brought into contact with the bath. Here we consider qubit probes, hence $d=2$. During the interaction, information
about the bath temperature is encoded into the quantum state of the probe through
temperature-dependent system-bath dynamics. After an interaction time $t$, the
$\beta$-encoded probe state $\rho_S(\beta,t)$ is detached from the bath and
measured in order to infer the value of $\beta$.

In the single-shot estimation scenario, the precision of unbiased estimation of
$\beta$ is bounded by the quantum Cramér–Rao inequality~\cite{SLD1,Braunstein1,holevo,Metrev2},
\begin{equation}
\Delta^2\beta \ge \frac{1}{\mathcal{F}(\beta)},
\end{equation}
where $\mathcal{F}(\beta)$ denotes the quantum Fisher information (QFI) associated
with the probe state $\rho_S(\beta,t)$. For $N$ independent repetitions of the
measurement, the bound improves to $\Delta^2\beta \ge 1/[N\,\mathcal{F}(\beta)]$.
Throughout this work, however, we focus exclusively on the single-shot estimation
regime ($N=1$).

The QFI quantifies the ultimate sensitivity of the probe state to variations in
the parameter $\beta$ and is defined as
\begin{equation}
\mathcal{F}(\beta)
= \Tr\!\left[\rho_S(\beta,t)\,L_\beta^2\right],
\label{f}
\end{equation}
where $L_\beta$ denotes the symmetric logarithmic derivative (SLD)~\cite{SLD1,Braunstein1}. The SLD
is implicitly defined through the operator equation
\begin{equation}
\partial_\beta \rho_S(\beta,t)
= \frac{1}{2}\!\left(\rho_S(\beta,t)\,L_\beta
+ L_\beta\,\rho_S(\beta,t)\right).
\end{equation}
In the eigenbasis $\{\ket{\tilde{i}}\}$, with $i=0,1$, of the probe state
$\rho_S(\beta,t)$, the matrix elements of the SLD are given by
\begin{equation}
\mathcal{L}_{ij} =
\begin{cases}
\displaystyle
\frac{2\,\bra{\tilde{i}}\,\partial_\beta \rho_S(\beta,t)\,\ket{\tilde{j}}}
{e_i+e_j}, & e_i+e_j \neq 0, \\[2ex]
0, & e_i+e_j = 0,
\end{cases}
\end{equation}
where $\{e_i\}$ denote the eigenvalues of $\rho_S(\beta,t)$, corresponding to the eigenvector $\ket{\tilde{i}}$.

Thus, the QFI is given by
\begin{equation*}
\mathcal{F}_{\beta,t} = \sum_{i,j} \frac{2\,\big|\bra{\tilde{i}}\,\partial_\beta \rho_S(\beta,t)\,\ket{\tilde{j}}\big|^2}{e_i + e_j}, \quad \text{for } e_i + e_j \neq 0.
\label{QFM}
\end{equation*}

The dependence of the probe state $\rho_S(\beta,t)$ on $\beta$ arises from its
interaction with the thermal bath. This encoding of
$\beta$ into the probe state enables its estimation via suitable quantum
measurements. When the measurement corresponds to the eigenbasis of the SLD, the
quantum Cramér–Rao bound can, in principle, be saturated.

For a given probe Hamiltonian $H_S$, conventional thermometric studies often consider the thermal steady state $\tau_S = \frac{e^{-\beta H_S}}{\Tr[e^{-\beta H_S}]}$ of the probe, for which the QFI reduces to $\mathcal{F}(\beta) = \mathrm{Var}_{\tau_S}(H_S)$. Here, $\mathrm{Var}_{\tau_S}(H_S) = \Tr[H_S^2 \tau_S] - \big(\Tr[H_S \tau_S]\big)^2$ denotes the variance of $H_S$ evaluated with respect to $\tau_S$.

However, several studies~\cite{cofi,ferm} have demonstrated that non-equilibrium probe states can lead to an enhancement of the QFI beyond the thermal-state limit. Motivated by these results, the present work focuses on \emph{transient quantum thermometry}, where parameter estimation is performed at finite times during the probe’s evolution, prior to reaching the steady or thermal state. We investigate this in both Markovian and non-Markovian system–bath interaction regimes. The Markovian case is introduced in the following section.
 
\subsection{The set-up for Markovian dynamics}
In this section, we discuss the basic setup for quantum thermometry under Markovian dynamics. In this setting, a quantum probe $S$ is weakly coupled to a bosonic bath $B$, whose temperature $T$ is to be estimated. The Hamiltonian of the probe is given by
\begin{equation}
H_S = \omega \sigma_z,
\end{equation}
with eigenvectors $\ket{0}$ and $\ket{1}$, corresponding to the eigenvalues $\omega_0=-\omega$ and $\omega_1=\omega$, respectively.

The bath Hamiltonian is
\begin{equation}
H_B = \int_{0}^{\omega_c} a_{\tilde{\omega}}^\dagger a_{\tilde{\omega}} \, d\tilde{\omega}.
\end{equation}
Here, $a_{\tilde{\omega}}$ and $a_{\tilde{\omega}}^\dagger$ denote the bath annihilation and creation operators corresponding to the frequency mode $\tilde{\omega}$.

The interaction between the probe and the bath is described by the interaction Hamiltonian
\begin{equation}
H_I = \int_{0}^{\omega_c} h(\tilde{\omega}) \left( \sigma_{-} a_{\tilde{\omega}}^\dagger + \sigma_{+} a_{\tilde{\omega}} \right) d\tilde{\omega}.
\end{equation}
Here, $\omega_c$ is the bath cut-off frequency, and $h(\tilde{\omega})$ denotes the coupling strength between the probe and the bath for a given frequency mode $\tilde{\omega}$. The operators $\sigma_- = \ketbra{0}{1}$ and $\sigma_+ = \ketbra{1}{0}$ are the lowering and raising operators of the probe, respectively.

The total Hamiltonian thus reads
\begin{equation}
H=H_S+H_B+H_I.
\end{equation}
The joint system-bath evolution is governed by the unitary operator
\begin{equation}
U_{SB}(t)=e^{-iHt}.
\end{equation}

We assume that initially the probe is prepared in a state $\rho_S(0)$, while the bath is in a thermal Gibbs state
\begin{equation}
\tau_B=\frac{e^{-H_B/T}}{\Tr\left[e^{-H_B/T}\right]},
\end{equation}
corresponding to the temperature $T$. Note that throughout the manuscript, we work in natural units with $\hbar = k_B = 1$, where $\hbar$ denotes Planck’s constant and $k_B$ is the Boltzmann constant. After a time $t$, the composite system evolves as
\begin{equation}
\rho_{SB}(t)=U_{SB}(t)\,\rho_S(0)\otimes\tau_B\,U_{SB}^\dagger(t).
\end{equation}

Since the probe is weakly coupled to a large bosonic environment, a series of standard approximations is invoked to derive an effective equation of motion for the reduced state of the probe. First, under the \emph{Born approximation}, the system-bath coupling is assumed to be sufficiently weak such that system-bath correlations remain negligible throughout the evolution, and the joint state can be approximated as
\begin{equation}
\rho_{SB}(t)\approx\rho_S(t)\otimes\tau_B.
\label{cp}
\end{equation}
Second, the \emph{Markov approximation} is justified by assuming that the bath relaxation (or correlation) time $t_R$ is much shorter than the characteristic relaxation time of the probe, $t_S$. This separation of time scales,
\begin{equation}
t_R \ll t_S,
\label{inf}
\end{equation}
ensures that the bath retains no memory of its past interactions with the probe. 
%and there can be no information backflow into the system.

Finally, the \emph{secular (rotating-wave) approximation}, is employed, which neglects rapidly oscillating terms that couple different Bohr frequencies of the system Hamiltonian. For a two-level probe with transition frequency $\omega_{01}=\omega_1-\omega_0$ and $\omega_{10}=\omega_0-\omega_1$, this approximation is valid when
\begin{equation}
|\omega_{01}-\omega_{10}|^{-1} \ll t_S,
\label{sec}
\end{equation}
ensuring that different transition channels evolve independently and that populations and coherences associated with distinct Bohr frequencies do not mix.

These three approximations together ensure that the resulting dynamics of the probe is CP-divisible~\cite{RHP,CPrev1,CPrev2}. This, in turn, guarantees that the distinguishability between any pair of system states decreases monotonically during the evolution. More precisely, let $\rho_1$ and $\rho_2$ be two distinct initial preparations of the probe, and let $\rho_1(t)$ and $\rho_2(t)$ denote their corresponding time-evolved states. Then, under CP-divisible dynamics, it is guaranteed that
\begin{equation}
D\big(\rho_1(t), \rho_2(t)\big) \leq D\big(\rho_1, \rho_2\big),
\end{equation}
where $D(\rho_1,\rho_2)$ denotes the distinguishability between the two states. A commonly used measure of distinguishability is the trace distance, defined as
\begin{equation}
D(\rho_1,\rho_2) = \frac{1}{2}\|\rho_1 - \rho_2\|_1,
\end{equation}
where the trace norm of an operator $A$ is given by $\|A\|_1 = \mathrm{Tr}\big[\sqrt{A^\dagger A}\big].$ This monotonic decrease in distinguishability is considered a signature of unidirectional information flow from the system to its environment, and is therefore indicative of the absence of information backflow, in Markovian dynamics.

Under the combined Born, Markov, and secular approximations, the reduced dynamics of the probe is governed by Gorini-Kossakowski-Sudarshan-Lindblad (GKSL) master equation~\cite{gksl1,gksl2,RH,gksl3},
\begin{equation}
\frac{d\rho_S(\beta,t)}{dt}
=-i[H_S,\rho_S(t)]+\mathcal{L}\big(\rho_S(t)\big),
\label{g}
\end{equation}
Here, $\mathcal{L}$ denotes the Lindblad dissipator determined by the bath
spectral density and temperature, and is given by
\begin{equation}
\begin{split}
\mathcal{L}(\rho)
=&\;\Gamma_{\downarrow}\!\left(\sigma_- \rho \sigma_+
- \frac{1}{2}\{\sigma_+\sigma_-,\rho\}\right) \\
&+ \Gamma_{\uparrow}\!\left(\sigma_+ \rho \sigma_-
- \frac{1}{2}\{\sigma_-\sigma_+,\rho\}\right).
\end{split}
\end{equation}
Here, $\{P,Q\}=PQ+QP$ denotes the anticommutator between two operators $P$ and $Q$.

The transition rates are given by
\begin{equation}
\Gamma_{\downarrow}
= J(\omega_{01})\bigl[1+\eta(\omega_{01})\bigr],
\qquad
\Gamma_{\uparrow}
= J(\omega_{01})\,\eta(\omega_{01}),
\end{equation}
where
\begin{equation}
\eta(\omega_{01})=\frac{1}{e^{\omega_{01}/T}-1}
\end{equation}
is the Bose-Einstein distribution function, and $J(\omega_{01})$ is the bath
spectral density evaluated at the system transition frequency $\omega_{01}$.

Throughout this manuscript, we consider an Ohmic spectral density, for which
\begin{equation}
J(\omega_{01})=\kappa\,\omega_{01}\equiv\gamma,
\label{gamma}
\end{equation}
where $\kappa$ is the proportionality constant characterizing the
system-bath coupling strength. The characteristic timescale, $t_S$ of the system
dynamics is set by $(\Gamma_{\uparrow}+\Gamma_{\downarrow})^{-1}$. Accordingly, we assume $\kappa$ to be sufficiently small so as to satisfy the
condition given in Eq.~\ref{sec} and to remain within the weak-coupling limit.

For a two-level system, the GKSL master equation~\eqref{g} can be solved
explicitly to obtain the probe state $\rho_S(\beta,t)$ at time $t$; the closed-form
expression is provided in the subsequent section~\ref{mar}. Initially, the probe does not
interact with the bath and therefore contains no information about the
temperature $T$. As the system evolves under the system-bath interaction, it
gradually acquires information about the bath, rendering the state
$\rho_S(\beta,t)$ a  function of $\beta$ and $t$. Consequently, one can compute the QFI with respect to the temperature $T$ using the
probe state $\rho_S(\beta,t)$ and the formula~\eqref{QFM}.

Having presented the basic set-up for Markovian quantum thermometry, we next
introduce the framework for non-Markovian dynamics, obtained by relaxing the
condition of no information backflow into the system. The set-up is as follows.

\subsection{The set-up for non Markovian dynamics}
In this section, we briefly discuss the two types of non-Markovian settings considered in this study: 

(i) Non-Markovian dynamics arising from the coupling of the system to an auxiliary degree of freedom, which is itself coupled to a thermal bath. 

(ii) Non-Markovian dynamics described within the framework of a quantum collisional model. 

Each of these settings is briefly discussed below.

\subsubsection{Auxilary assisted non-Markovianity}
\label{ANM}

Recall that a Markovian description of the probe dynamics relies on the absence of information backflow from the environment to the system. One way to  effectively induce information backflow and thereby introduce non-Markovianity into the dynamics of the probe is to
bring an auxiliary $A$.

The auxiliary acts an intermediate subsystem and mediates the interaction between the probe
$S$ and the thermal bath $B$, with inverse temperature $\beta$. 
%Our goal is to estimate $\beta$. 
The probe and auxiliary interact coherently, while
only the auxiliary is directly coupled to the bath. We consider a qubit auxiliary. The Hamiltonian of the
composite probe-auxiliary system $S\!A$ is given by
\begin{equation}
H_{SA}
=\omega\,\sigma_z\otimes\mathbbm{I}
+\omega_A\,\mathbbm{I}\otimes\sigma_z
+J\bigl(\sigma_{+}^S\otimes\sigma_{-}^A
+\sigma_{-}^S\otimes\sigma_{+}^A\bigr),
\end{equation}

where, $\sigma_{\pm}^{S/A}$ denote the raising and lowering (creation and
annihilation) operators acting on the probe and auxiliary systems,
respectively. $J$ denotes the coherent coupling strength. And  the local field strengths $\omega=\omega_A.$

We assume that the introduction of the auxiliary does not compromise the Markovian character of the global dynamics of the probe-auxiliary system, which remains governed by a GKSL master equation. However, the coherent $S\!A$ interaction generates correlations between the probe and the auxiliary during the evolution.

As a consequence, the auxiliary effectively acts as an environment for the probe with which it becomes correlated. When the auxiliary is traced out, these system–auxiliary correlations influence the reduced dynamics of the probe. More specifically, the presence of correlations leads to  increase in the distinguishability between pairs of probe states, even though the joint $S\!A$ evolution itself remains Markovian. This increase in distinguishability is a signal of information backflow into the system and the resulting non-Markovian behavior can be quantified using the Breuer–Laine–Piilo (BLP) measure~\cite{BLP}(discussed in detail in Sec.~\ref{non-mar}),

We emphasize that this apparent information backflow does not originate from memory effects in the external bath, but rather from tracing out part of a larger Markovian system in which correlations are dynamically generated.

During the evolution, information about the bath inverse temperature $\beta$ is
first encoded in the state of the auxiliary and subsequently transferred to the
probe via the coherent $S\!A$ interaction, giving rise to a non-Markovian
dependence of the probe state on $\beta$. Similar non-Markovian effects induced
by fermionic baths have been studied in Ref.~\cite{ferm}.

The GKSL master equation governing the evolution of the composite probe-auxiliary system is given as 
\begin{equation}
    \frac{d\rho_{SA}(\beta,t)}{dt}
=-i[H_{SA},\rho_{SA}(t)]+\mathcal{L}\big(\rho_{SA}(t)\big),
\label{g2}
\end{equation}
where 
\begin{equation}
\begin{split}
\mathcal{L}(\rho_{SA})
=&\;\Gamma_{\downarrow}\!\left(\sigma_-^A \rho_{SA} \sigma_+^A
- \frac{1}{2}\{\sigma_+^A\sigma_-^A,\rho_{SA}\}\right) \\
&+ \Gamma_{\uparrow}\!\left(\sigma_+^A \rho_{SA} \sigma_-^A
- \frac{1}{2}\{\sigma^A_-\sigma^A_+,\rho_{SA}\}\right).
\end{split}
\end{equation}
Thus the reduced state of the probe at transient time $t$, is given as $\rho_S(\beta,t)=\Tr_A[\rho_{SA}(\beta,t)]$. In sec.~\ref{non-mar}, we present the exact analytical form of $\rho_S(\beta,t)$, and calculate the QFI, corresponding to the inverse temperature $\beta$.

In the subsection below we briefly discuss the quantum collisional model for non-Markovianity.

\subsubsection{The quantum collisional model for non-Markovianity}
\label{qc}
The system-bath interaction in open quantum dynamics can be modeled by a repeated interaction scheme known as the quantum collisional model~\cite{Colmem,ColmemFt}, whereby the environment is represented as a large collection of identical and initially uncorrelated ancillas. Let us denote these ancillas by $A_i$, where $i=1,2,\ldots,\infty$ labels the $i^{\text{th}}$ ancilla. The evolution of the probe system $S$ is governed by successive collisions between the system and the ancillas, occurring one at a time.

Between two successive collisions of the system with ancillas $A_i$ and $A_{i+1}$, the ancilla $A_i$ may interact with $A_{i+1}$. Such ancilla-ancilla interactions introduce memory effects or non-Markovianity into the dynamics. This is in contrast to the Markovian collision model case, where the bath retains no memory of the past evolution of the system, corresponding to the absence of ancilla-ancilla interactions.

The collision between the system and the $i^{\text{th}}$ ancilla is described by the unitary operator
\begin{equation}
U_{SA_i}=e^{-iH_{SA_i}t_c},
\end{equation}
where $t_c$ is the interaction time and $H_{SA_i}$ is the interaction Hamiltonian between the system and the $i^{\text{th}}$ ancilla. Let $\sigma$ represent the joint state of the probe and the environment. The system–ancilla collision then induces the transformation
\begin{equation}
\sigma \rightarrow U_{SA_i}\sigma U_{SA_i}^{\dagger}.
\end{equation}

The interaction between successive ancillas is modeled by a probabilistic swap operation $\mathcal{S}_{i,i+1}$ acting as
\begin{equation}
\mathcal{S}_{i+1,i}(\sigma)=(1-P)\sigma + P\,\hat{S}_{i+1,i}\sigma\hat{S}_{i+1,i}^{\dagger},
\end{equation}
where $\hat{S}_{i+1,i}$ is the swap operator acting on ancillas $A_{i+1}$ and $A_i$, and $P$ denotes the swapping probability.

Consequently, after the $n^{\text{th}}$ collision, the state of the composite system can be written as
\begin{equation}
\sigma =
U_{SA_n} \circ \mathcal{S}_{n-1,n} \circ \cdots \circ
U_{SA_2} \circ \mathcal{S}_{1,2} \circ U_{SA_1}(\sigma_0).
\end{equation}

Ref.~\cite{Colmem} showed that in the continuous-time limit, where the duration of each system--ancilla collision $t_c \to 0$ and the number of collisions $n \to \infty$, the reduced dynamics of the system can be described by the master equation
\begin{equation}
\frac{d\rho(t)}{dt}=
\Gamma \int_{0}^{t} dt' \, \mathcal{E}_t' 
\Big(\frac{d\rho(t-t')}{dt}\Big)
+ e^{-\Gamma t}\frac{d\mathcal{E}_t(\rho_0)}{dt}.
\end{equation}

Here $\Gamma$ characterizes the strength of memory effects arising from ancilla-ancilla interactions and is related to the swapping probability through
\begin{equation}
P = e^{-\Gamma t_c}.
\end{equation}
The limit $\Gamma=0$ corresponds to perfect swapping ($P=1$) between successive ancillas, whereas $\Gamma \to \infty$ corresponds to the absence of ancilla--ancilla interactions ($P=0$), resulting in Markovian dynamics. In this work, we focus on the regime of strong non-Markovianity and therefore consider $\Gamma=0$, corresponding to $P=1$.

As can be seen from the above master equation, for $\Gamma=0$ the system state at time $t$ can be expressed as
\begin{equation}
\rho(t)=\mathcal{E}_t(\rho_0).
\end{equation}
  The explicit form of the dynamical map $\mathcal{E}_t(\cdot)$ depends on the specific system–ancilla interaction. In this work, we consider a Jaynes–Cummings interaction between the system and the ancillas. Furthermore, the bath ancillas are taken to be qubits.

It is worth noting that the collisional model considered in this work differs significantly from existing approaches to collisional quantum thermometry~\cite{Colth1,Colth2,Colth3,Colth4}. In typical schemes, the bath temperature is estimated through repeated interactions between a system and a sequence of identically prepared auxiliary units, where the system itself is coupled to the bath and acts as an intermediary. After many such collisions, the joint auxiliary state is measured to infer the temperature. In contrast, our setup does not involve such a mediating system. Instead, the bath itself is modeled as a collection of many non-interacting, identically prepared auxiliaries at thermal equilibrium that sequentially interact directly with the probe. An auxiliary that has previously interacted with the probe  may interact with the next incoming auxiliary  before the latter collides with the probe. This mechanism effectively incorporates environmental memory into the dynamics. After a large number of collision the probe is measured to estimate the temperature.

Before proceeding to the non-Markovian dynamics, we first analyze the Markovian scenario in the next section. We show that, under Markovian dynamics, initially colder probes are both necessary and sufficient to achieve enhanced precision at transient times. The detailed analysis is presented below.
\section{Colder probe bias under Markovian dynamics}
\begin{figure*}
\centering
\vspace{-2cm}
\includegraphics[scale=0.26]{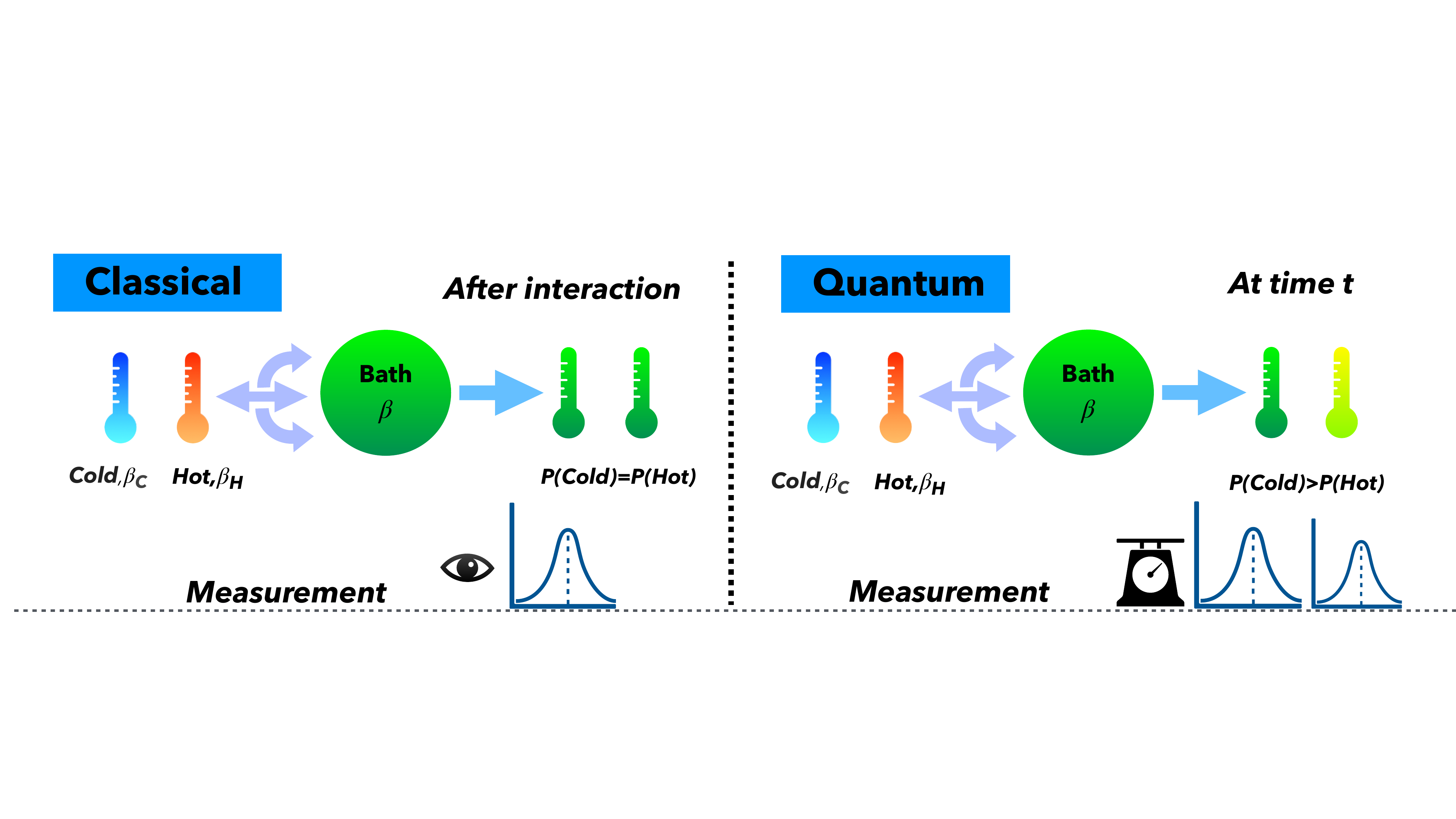}
\vspace{-2cm}
\caption{\textbf{Analogy between classical and quantum thermometry.} The figure illustrates two scenarios classical thermometry on the left and quantum thermometry on the right. In the classical case both initially cold and initially hot thermometers after interacting with the thermal bath for a sufficiently long time equilibrate with the bath and achieve the same maximal precision in temperature estimation. Thus the long time precision is independent of the initial state of the thermometer. In contrast in the quantum case the probe and bath dynamics leads to a strong dependence of the transient estimation precision on the initial state of the probe. In particular probes prepared at a temperature lower than that of the bath can yield enhanced precision at finite times whereas initially hotter probes never surpass the precision attainable by the thermal state throughout the evolution. This highlights a genuine quantum bias in the choice of the initial probe state for achieving enhanced precision at transient times.}
	\label{Pic1}
\end{figure*} 
\label{mar}
We are interested in understanding how the initial temperature of the probe
affects the estimation precision at transient times during the evolution.
Recall that the probe Hamiltonian is given by $H_S=\omega\sigma_z$, with
eigenenergies $-\omega$ and $\omega$ corresponding to the eigenvectors
$\ket{0}$ and $\ket{1}$, respectively. The temperature of an initial state can
be unambiguously defined when the state is diagonal in the energy eigenbasis.
Motivated by this, we restrict ourselves to initial states of the form
\begin{equation}
\rho_i = (1-p)\ketbra{0}{0} + p\ketbra{1}{1}.
\label{16}
\end{equation}
For such a state, the corresponding inverse temperature $\beta_i$ is given by
\begin{equation}
\beta_i = \frac{1}{2\omega} \ln\!\left(\frac{1-p}{p}\right).
\end{equation}

We further restrict our analysis to $\beta_i \geq 0$, which imposes the
constraint $0 \leq p \leq 1/2$. In other words, we exclude states exhibiting
population inversion or negative temperature. Within this range, one always
has $p \leq 1-p$.

 The thermal state $\tau_S$ corresponding to $H_S$ at inverse temperature $\beta$ is a
special case of the state $\rho_i$, with
\begin{equation}
p = p^{e} =
\frac{e^{-\omega\beta}}{e^{\omega\beta}+e^{-\omega\beta}}.
\label{therm}
\end{equation}
Here, the superscript $e$ in $p^{e}$ denotes equilibrium.
Consequently, $\beta_i < \beta$ implies $p > p^{e}$, indicating that
the initial state is hotter than the thermal state $\tau_S$. Conversely,
$p < p^{e}$ implies $\beta_i > \beta$, meaning that the initial state
is colder than $\tau_S$.

Having specified the form of the initial state, we now turn to the solution of
the probe state at a transient time $t$, governed by the GKSL master equation.
As shown in Ref.~\cite{cofi}, the solution of the master equation yields
\begin{equation}
\rho_S(\beta,t) = \bigl(1-e^{\lambda t}\bigr)\tau_S + e^{\lambda t}\rho_i,
\label{rf}
\end{equation}
where
\begin{equation}
\lambda = \frac{\gamma}{2p^{e}-1}.
\label{lambda}
\end{equation}

Since we have $0 \leq p \leq 1/2$ and $\gamma > 0$, we always have $\lambda < 0$.
In the long-time limit, the system relaxes to the thermal state $\tau_S$. The state $\rho_S(\beta,t)$ is therefore a function of the initial inverse temperature $\beta$, since both $\tau_S$ and $p^{\mathrm{th}}$ depend on $\beta$, as well as a function of time.

Once the state at time $t$ is obtained, one can compute the QFI. Let $\mathcal{F}(\beta,t)$ denote the QFI corresponding to
the state $\rho_S(\beta,t)$, and let $\mathcal{F}(\beta)$ denote the QFI of
the thermal state $\tau_S$. In what follows, we present a theorem stating that,
at a transient time $t$, the condition $\mathcal{F}(\beta,t) >
\mathcal{F}(\beta)$ is possible if and only if the initial state satisfies
$\beta_i > \beta$. In other words, hotter initial states can never lead to an
enhancement in precision at transient times. This bias is purely a quantum phenomenon, because in classical thermometry, the bath temperature can ultimately be determined with unit precision regardless of the initial temperature of the thermometer, provided sufficient time has elapsed. In Fig.~\ref{Pic1}, we pictorially illustrate the analogy between classical and quantum thermometry. The theorem is stated as follows.

\begin{theorem}
Let the initial probe state $\rho_i$, of the form given in Eq.~\eqref{16}, with
inverse temperature $\beta_i=\frac{1}{2\omega}\ln\!\left(\frac{1-p}{p}\right)$,
evolve under Markovian dynamics while interacting with a thermal bath at inverse
temperature $\beta \in(0,\infty)$. Let $\rho_S(\beta,t)$ denote the probe state at time $t$,
given by Eq.~\eqref{rf}, and used to estimate $\beta$. Then, there exists a
finite time $t>0$ such that the transient QFI
$\mathcal{F}(\beta,t)$ exceeds the steady-state QFI $\mathcal{F}(\beta)$ associated with the thermal state $\tau_S$ if and only if
$\beta_i>\beta$.
\end{theorem}

\begin{proof}
We have
\begin{equation*}
\begin{split}
\rho_S(\beta,t)
&= \bigl(1-e^{-\lambda t}\bigr)\tau_S + e^{-\lambda t}\rho_i \\
&= \bigl[1-q(\beta,t)\bigr]\ketbra{0}{0}
  + q(\beta,t)\ketbra{1}{1}.
\end{split}
\end{equation*}
where the population $q(\beta,t)$ is a function of both the inverse temperature
$\beta$ and the evolution time $t$. Using this expression, one can compute the QFI at a transient time $t$ as
\begin{equation}
\mathcal{F}(\beta,t)=\Tr\!\left[L_\beta^2\,\rho_S(\beta,t)\right].
\end{equation}

Here, $L_{\beta}$ denotes the SLD operator. Note that throughout the evolution the state $\rho_S(\beta,t)$ remains diagonal
in the energy eigenbasis. Consequently, the eigenbasis of $\rho_S(\beta,t)$
coincides with the eigenbasis of the system Hamiltonian $H_S$, with eigenvalues
$e_0 = 1-q(\beta,t)$ and $e_1 =q(\beta,t)$. Thus, the operator $L_{\beta}^2$ takes the form
\begin{equation*}
L_{\beta}^2
= \frac{\bigl(\partial_\beta q(\beta,t)\bigr)^2}{[1-q(\beta,t)]^2}\,\ketbra{0}{0}
+ \frac{\bigl(\partial_\beta q(\beta,t)\bigr)^2}{q(\beta,t)^2}\,\ketbra{1}{1}.
\end{equation*}
And we have 
\begin{equation}
\mathcal{F}(\beta,t)
= \frac{\bigl(\partial_\beta q(\beta,t)\bigr)^2}
       {q(\beta,t)\,[1-q(\beta,t)]}.
\end{equation}
By straightforward algebra, one finds that
\begin{equation*}
\partial_\beta q(\beta,t)
= 2\omega\, p^{e}\,[1- p^{e}]\,\delta(\beta,t),
\end{equation*}
where
\begin{equation*}
\delta(\beta,t)
= 1-e^{\lambda t}+2t\,(p^e-p)\,\frac{\lambda^2}{\gamma}e^{\lambda t}.
\end{equation*}
Here, $\lambda$ and $\gamma$ are given by Eqs.~\eqref{lambda} and~\eqref{gamma},
respectively. Using this expression, the quantum Fisher information takes the form
\begin{equation}
\mathcal{F}(\beta,t)
= \frac{A\,\delta(\beta,t)^2}{q(\beta,t)\,[1-q(\beta,t)]},
\end{equation}
with $A=4\omega^2\{p^{e}\,[1- p^{e}]\}^2$.
Also by the same method it is easy to check the QFI corresponding to the thermal state $\tau_S$, is given as 
\begin{equation*}
    \mathcal{F}(\beta)=4\omega^2\{p^{e}\,[1- p^{e}]\}.
\end{equation*}
Thus we have the ratio
\begin{equation*}
    r=\frac{\mathcal{F}(\beta,t)}{\mathcal{F}(\beta)}=\frac{\delta(\beta,t)^2p^{e}\,[1- p^{e}]}{q(\beta,t)\,[1-q(\beta,t)]}.
\end{equation*}
Using the expression for $\rho_S(\beta,t)$, one can show that
\begin{equation*}
q(\beta,t)\,[1-q(\beta,t)]
= p^{e}\,[1-p^{e}]\;\alpha(\beta,t),
\end{equation*}
where
\begin{equation*}
\begin{split}
\alpha(\beta,t)
&= (1-e^{\lambda t})^2
+ e^{\lambda t}(1-e^{\lambda t})
\left(\frac{p}{p^{e}}+\frac{1-p}{1-p^{e}}\right) \\
&\quad
+ e^{2\lambda t}\frac{p\,[1-p]}{p^{e}\,[1-p^{e}]}.
\end{split}
\end{equation*}
Therefore, we define
\begin{equation*}
r=\frac{\delta(\beta,t)^2}{\alpha(\beta,t)}.
\end{equation*}
At a given time $t>0$, if $r>1$, then $\mathcal{F}(\beta,t)>\mathcal{F}(\beta)$.
We now show that the condition $\beta_i>\beta$ is both necessary and sufficient
for the existence of a finite time $t>0$ at which $r>1$.
The proof is divided into two parts:
(i) we first show that $\beta_i>\beta$ is necessary for $r>1$ to occur at some
$t>0$, and
(ii) we then show that $\beta_i>\beta$ is sufficient to ensure the existence of
a time $t>0$ such that $r>1$.

\medskip
\noindent\textit{(i) Cold probes are necessary for a transient enhancement in precision.}
We begin by considering hotter initial probe states, for which
$\beta_i<\beta$. In this case, we show that both $\delta(\beta,t)\leq 1$ and
$\alpha(\beta,t)^{-1}\leq 1$ for all $t\geq 0$. Consequently,
$r\leq 1$ at all times, implying that a transient enhancement in precision is
not possible with hotter probes.

To this end, the function $\alpha(\beta,t)$ can be equivalently written as
\begin{equation*}
\begin{split}
\alpha(\beta,t)
&= 1 - e^{2\lambda t}\!\left[\frac{(p^{e}-p)^2}{p^{e}(1-p^{e})}\right]
+ e^{\lambda t}\frac{(p^{e}-p)(2p^{e}-1)}{p^{e}(1-p^{e})} \\
&= 1 - a\,e^{2\lambda t} + b\,e^{\lambda t},
\end{split}
\end{equation*}
where
$a=\frac{(p^{e}-p)^2}{p^{e}(1-p^{e})}>0$ and
$b=\frac{(p^{e}-p)(2p^{e}-1)}{p^{e}(1-p^{e})}>0$(Since for hotter initial state $p^{e}<p$ and $2p^{e}-1<0$).

To determine the extrema of $\alpha(\beta,t)$ with respect to time, for a fixed
$\beta$, we analyze its behavior over the domain $t\in[0,\infty)$. Since
$\alpha(\beta,t)$ is a smooth function of $t$, any extremum must occur either at
the boundaries ($t=0$ or $t\to\infty$) or at a stationary point satisfying
$\partial_t\alpha(\beta,t)=0$.

 We first evaluate $\alpha(\beta,t)$ at $t=0$. At $t=0$, one
finds
\begin{equation*}
\begin{split}
\alpha(\beta,0)
&= 1-\frac{(p^{e}-p)^2}{p^{e}(1-p^{e})}
+ \frac{(p^{e}-p)(2p^{e}-1)}{p^{e}(1-p^{e})}\\
&= 1+\frac{(p-p^{e})(1-p^{e}-p)}{p^{e}(1-p^{e})}.
\end{split}
\end{equation*}
For hotter initial states we have $p>p^{e}$. Moreover,
\begin{equation*}
1-p^{e}-p
=\frac{e^{\omega(\beta+\beta_i)}-e^{-\omega(\beta+\beta_i)}}{Z_e Z_i}
>0,
\end{equation*}
where $Z_e=e^{\omega\beta}+e^{-\omega\beta}$ and
$Z_i=e^{\omega\beta_i}+e^{-\omega\beta_i}$.
This implies that $\alpha(\beta,t)>1$ at $t=0$.

And at the initial time $t \to \infty$, we immediately obtain $\alpha(\beta,0)=1$.

Next, we examine the interior of the domain, $0<t<\infty$. Setting
$\partial_t\alpha(\beta,t)=0$, we obtain the stationary point
\begin{equation*}
e^{\lambda t}=\frac{b}{2a}.
\end{equation*}
Evaluating the second derivative at this point yields
\begin{equation*}
\frac{\partial^2\alpha(\beta,t)}{\partial t^2}
=-\lambda^2 e^{\lambda t}<0,
\end{equation*}
which shows that this stationary point corresponds to a \emph{maximum} of
$\alpha(\beta,t)$. The corresponding maximal value is
$\alpha(\beta,t)=1+\frac{b^2}{4a}$.

Since the only stationary point in the interval $0<t<\infty$ is a maximum, the
minimum of $\alpha(\beta,t)$ must occur at the boundaries. As
$\alpha(\beta,0)>1$ and $\alpha(\beta,t\to\infty)=1$, the minimum is attained
at $t \to \infty$, with $\alpha_{\min}=1$. Therefore, for hotter initial states,
$\alpha(\beta,t)\geq 1$ for all $t\geq 0$, which implies
$\alpha(\beta,t)^{-1}\leq 1$.

We now turn to the behavior of $\delta(\beta,t)$. As in the case of
$\alpha(\beta,t)$, the extrema of $\delta(\beta,t)$ can occur either at the
boundary points $t=0$ and $t\to\infty$, or at interior stationary points
satisfying $\partial_t \delta(\beta,t)=0$.

We first examine the boundary values. In the long-time limit,
\begin{equation*}
\lim_{t\to\infty} \delta(\beta,t)=1,
\end{equation*}
whereas at the initial time we have
\begin{equation*}
\delta(\beta,0)=0.
\end{equation*}

We now analyze the interior of the domain $t\in(0,\infty)$. Solving the
stationary condition $\partial_t \delta(\beta,t)=0$ yields a critical point at
\begin{equation*}
t=t_h=\frac{\lambda-m}{\lambda m},
\end{equation*}
where
\begin{equation*}
m=2\,(p^{e}-p)\,\frac{\lambda^2}{\gamma}.
\end{equation*}
For hotter initial states, characterized by $p>p^{e}$, we have $m< 0$.

Evaluating the second derivative at the stationary point gives
\begin{equation*}
\frac{\partial^2 \delta(\beta,t)}{\partial t^2}
=m\lambda e^{\lambda t}>0,
\end{equation*}
which shows that $t=t_h$ corresponds to a minimum of $\delta(\beta,t)$.
The value of $\delta(\beta,t)$ at this point is
\begin{equation*}
\delta(\beta,t_h)
=1-\left|\frac{m}{\lambda}\right|e^{\lambda t_h}.
\end{equation*}

The quantity $\delta(\beta,t_h)$ could become negative only if
$\left|m/\lambda\right|>1$. However, for hotter initial states one finds
\begin{equation*}
|m|-|\lambda|
=\frac{\gamma(2p-1)}{(2p^{e}-1)^2}\le 0,
\end{equation*}
since $0\le p\le 1/2$.
Thus we have $t_h\leq0$. Since time cannot be negative this shows no extrema occurs at the interior.

Combining the boundary and interior analyses, we conclude that for all
$t\in[0,\infty)$,
\begin{equation*}
0\le \delta(\beta,t)\le 1.
\end{equation*}
Consequently,
\begin{equation*}
0\le \delta(\beta,t)^2\le 1.
\end{equation*}

Since we have already shown that $\alpha(\beta,t)^{-1}\le 1$ for hotter initial
states, it follows that
\begin{equation*}
r=\frac{\delta(\beta,t)^2}{\alpha(\beta,t)}\le 1
\quad \text{for all } t\ge 0.
\end{equation*}
Therefore, for hotter initial probes, the quantum Fisher information cannot
exceed its steady-state value at any transient time, and no enhancement in
precision is possible.

But for colder initial states we have. 
\begin{equation*}
\begin{split}
\alpha(\beta,t)
&= 1 - e^{2\lambda t}\!\left[\frac{(p^{e}-p)^2}{p^{e}(1-p^{e})}\right]
+ e^{\lambda t}\frac{(p^{e}-p)(2p^{e}-1)}{p^{e}(1-p^{e})} \\
&= 1 - a\,e^{2\lambda t} + b\,e^{\lambda t},
\end{split}
\end{equation*}
Notice for colder initial states we have $p^e>p$ and $2p^e-1 < 0$. Hence we have $b < 0$, and $a>0$. Thus $\alpha <1$, and $\alpha^{-1}>1$, for all $0\leq t<\infty$. At $t=\infty$ we have $\alpha=1$. Combining both we get, $\alpha^{-1}\geq1$ for $t\in[0,\infty)$. So if we can get $\delta^2(\beta,t)\geq 1$ for any $t$, this can make $r > 1$, ensuring enhancement. 

Now the boundary values of are still the same i.e $\delta(\beta,0)=0$ and $\delta(\beta,t\to \infty)=1$. However now within the interior $t \in(0,\infty)$, the extrema occurs at $t_c=\frac{\lambda-n}{n\lambda}>0$. Where $n=2(p^e-p)\frac{\lambda^2}{\gamma}>0$. Such that at $t_c$, $\partial_t\delta(\beta,t)=0$ and $\partial_t^2\delta(\beta,t)=n\lambda e^{\lambda t}<0$. This guarantees that $t_c$, is the point of maximum and the value of $\delta(\beta,t_c)$ is given as $$\delta(\beta,t_c)=1+\frac{n}{\abs{\lambda}}e^{\lambda t_c}.$$ Thus $$\delta(\beta,t_c)>1.$$ Therefore for $\beta_i>\beta$, one can have $r>1$. This proves that colder initial probes are necessary for transient enhancement in precision. Next we argue that colder initial probes are indeed sufficient for ensuring that the precision is enhanced at transient region. This can be seen as follows.

ii)\textit{Colder probes are sufficient for precision enhancement at transient times:} We have $\delta(\beta,t_c)>1$, and since $\alpha(\beta,t_c)>1$, we can consider the transient time $t_c$. The ratio 
\begin{equation}
r_c = \frac{\delta(\beta,t_c)^2}{\alpha(\beta,t_c)} > 1.
\end{equation}
The exact maximum of $r$ may occur at some other time $\tilde{t}$, which we denote by $r_{\max}$. By definition, we then have 
\begin{equation}
r_{\max} \geq r_c.
\end{equation}
Since the lower bound $r_c$ is already greater than 1, it follows that the actual maximum $r_{\max}$ must also be greater than 1. This demonstrates that estimating the bath's temperature with a colder probe is sufficient to achieve enhanced precision at a transient time during the probe–bath interaction. 

This ends the proof of Theorem 1.
\end{proof}

Thus, Theorem 1 reveals a bias in the initial probe temperature that is necessary and sufficient to attain enhanced precision at transient times. 

In the following subsections, we analyze how this bias is affected by the introduction of memory effects during the probe evolution, namely non-Markovian dynamics. Specifically, we consider two cases: (1) non-Markovianity arising from an auxiliary-mediated interaction with the bath, as discussed in Sec.~\ref{ANM}, and (2) a quantum collisional model, as discussed in Sec.~\ref{qc}.

\section{Bias Persistence under Auxiliary-Mediated Non-Markovianity}
\label{non-mar}

We consider the initial state of the composite probe-auxiliary system to be of the form 
$\rho_i \otimes \tau_{\beta}$, where $\rho_i$, characterized by an initial inverse temperature $\beta_i$, is diagonal in the energy eigenbasis and is given by Eq.~\eqref{16}. The auxiliary system is assumed to be in a thermal state at inverse temperature $\beta$, corresponding to the bath. 

The joint system evolves according to the GKSL master equation in Eq.~\eqref{g2}. The reduced state of the probe, $\rho_S(\beta,t)$, at time $t$ is obtained by solving this equation for the composite state $\rho_{SA}(\beta,t)$ and subsequently tracing out the auxiliary degrees of freedom.

To this end, we express $\rho_{SA}(\beta,t)$ in the eigenbasis of the total Hamiltonian 
$\omega(\sigma_z \otimes \mathbbm{I} + \mathbbm{I} \otimes \sigma_z)$, denoted by 
$\{\ket{lm}\}$, with $l,m \in \{0,1\}$. Defining the matrix elements 
$\rho_{lm,l'm'} \equiv \langle lm | \rho_{SA} | l'm' \rangle$ and their time derivatives 
$\dot{\rho}_{lm,l'm'}$, the time evolution of these elements follows from Eq.~\eqref{g2} as

\begin{equation} 
\begin{aligned}
\dot{\rho}_{00,00} &= \gamma_{\downarrow}\rho_{01,01} - \gamma_{\uparrow}\rho_{00,00}, \\ \dot{\rho}_{01,01} &= -iJ\bigl(\rho_{10,01}-\rho_{01,10}\bigr) + \gamma_{\downarrow}\rho_{01,01} + \gamma_{\uparrow}\rho_{00,00}, \\ \dot{\rho}_{11,11} &= -\gamma_{\downarrow}\rho_{11,11} + \gamma_{\uparrow}\rho_{10,10}, \\ \dot{\rho}_{10,10} &= +iJ\bigl(\rho_{10,01}-\rho_{01,10}\bigr) + \gamma_{\downarrow}\rho_{11,11} - \gamma_{\uparrow}\rho_{10,10}, \\ \dot{\rho}_{00,10} &= \dot{\rho}_{10,00} = 0, \\
\dot{\rho}_{01,11}&=\dot{\rho}_{01,11}=0,\\
\dot{\rho}_{10,11}&=\dot{\rho}_{11,10}=0,
\\ \dot{\rho}_{10,01} &= -iJ\bigl(\rho_{01,01}-\rho_{10,10}\bigr) - \Gamma_{T}\rho_{10,01}. 
\label{set}
\end{aligned} 
\end{equation} 

Here $\Gamma_T = \frac{\gamma_{\downarrow} + \gamma_{\uparrow}}{2}$ and for notational simplicity, we have suppressed the explicit dependence of the density matrix elements on $\beta$ and $t$. The above set of coupled equations allows one to determine $\rho_{SA}(\beta,t)$, and hence obtain the reduced probe state $\rho_S(\beta,t)$.

Therefore, the off-diagonal elements of the probe state satisfy
\begin{equation*}
    \dot{\rho}_S^{01} = \dot{\rho}_S^{10} = 0.
\end{equation*}
Since the probe is initially diagonal in the energy eigenbasis, it follows that it remains diagonal throughout the evolution.

The time evolution of the diagonal elements is given by
\begin{equation}
\begin{aligned}
\dot{\rho}_S^{00} &= \dot{\rho}_{00,00} + \dot{\rho}_{01,01} \\
&= -iJ\bigl(\rho_{10,01} - \rho_{01,10}\bigr), \\
\dot{\rho}_S^{11} &= -\dot{\rho}_S^{00}.
\end{aligned}
\end{equation}

Now let $\rho_{10,01} = x + i y$, and define $\Delta(t) = \rho_{01,01} - \rho_{10,10}$. Then we obtain
\begin{equation}
\dot{\rho}_S^{00} = 2J y,
\label{pop}
\end{equation}
and
\begin{equation}
\begin{aligned}
\dot{\Delta}(t) &= 4J y 
- \gamma_{\downarrow}\big(\rho_{01,01} + \rho_{11,11}\big) \\
&\quad + \gamma_{\uparrow}\big(\rho_{00,00} + \rho_{10,10}\big).
\end{aligned}
\label{delta}
\end{equation}

The quantities $\rho_{00,00} + \rho_{10,10}$ and $\rho_{01,01} + \rho_{11,11}$ correspond to the probabilities $p_0^A$ and $p_1^A$ of the auxiliary being in the ground and excited states of its local Hamiltonian, respectively. Hence, we have
\begin{equation}
    p_0^A = \frac{\dot{\Delta}(t) + \gamma_{\downarrow} - 4J y}{2\Gamma_T}.
\end{equation}
Furthermore, it follows from the expressions for $\dot{\rho}_{00,00}$ and $\dot{\rho}_{01,01}$ given in Eq.~\eqref{set} that
\begin{equation}
   \dot{p}_0^A = -2J y + \gamma_{\downarrow} p_1^A - \gamma_{\uparrow} p_0^A.
\end{equation}
In addition, the equation for $\dot{\rho}_{10,01}$ in Eq.~\eqref{set} yields
\begin{equation}
    \dot{y} = -J \Delta(t) - \Gamma_T y,
    \label{y}
\end{equation}
Using Eqs.~\eqref{delta}-\eqref{y}, we obtain
\begin{equation}
    \ddot{\Delta}(t) + 2\Gamma_T \dot{\Delta}(t) + 4J^2 \Delta(t) = 0.
\end{equation}

which is the equation of a damped harmonic oscillator. In the strong-coupling regime ($J \gg 2\Gamma_T$), relevant for non-Markovian memory effects, the dynamics corresponds to an underdamped oscillator with frequency $\Omega = \sqrt{4J^2 - \Gamma_T^2}$.

Imposing the initial conditions explicitly at $t=0$, namely $\Delta(0) = \Delta_0$ and $\dot{\Delta}(0) = 0$, the solution is given by
\begin{equation}
    \Delta(t) = \Delta_0\, e^{-\Gamma_T t} \left[ \cos(\Omega t) + \frac{\Gamma_T}{\Omega} \sin(\Omega t) \right].
\end{equation}

\begin{figure*}
\centering
\includegraphics[scale=0.8]{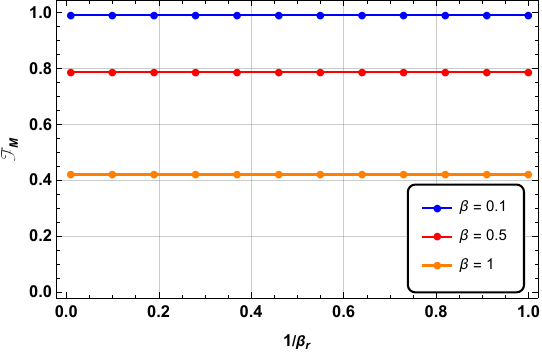}
\hspace{1cm}
\includegraphics[scale=0.8]{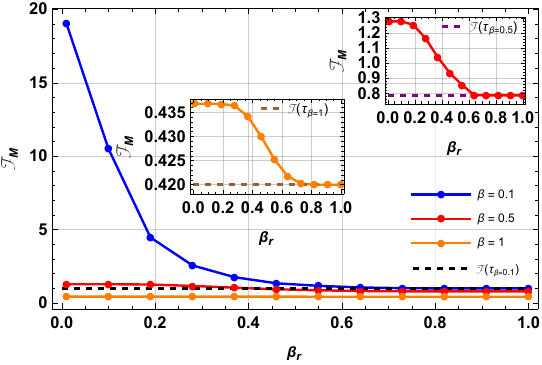}
\caption{\textbf{Persistence of cold probe bias under auxiliary-mediated interaction.}
The plot shows the maximum QFI, $\mathcal{F}_M$, optimized over time for initially hotter (left panel) and initially colder probes (right panel), for three values of the bath inverse temperature: $\beta = 0.1$ (blue), $\beta = 0.5$ (red), and $\beta = 1$ (orange). The system–auxiliary interaction strength is $J = 10$, and $\gamma = 10^{-4}$. As seen in the left panel, for initially hotter probes, the maximum QFI coincides with that of the thermal state corresponding to the same $\beta$, indicating the absence of any enhancement in precision at transient times. In contrast, for initially colder probes, a clear enhancement in precision is observed at transient times for all considered values of $\beta$. This enhancement is more pronounced for smaller values of $\beta_r$ and gradually diminishes as $\beta_r \to 1$, indicating that the colder the initial probe, the larger the value of $\mathcal{F}_M$ compared to the corresponding thermal state. The insets in the right panel show magnified views for $\beta = 0.5$ and $\beta = 1$. In the right panel, the values of $\mathcal{F}_M$ for the corresponding thermal states are shown as horizontal dashed lines: black for $\beta = 0.1$, purple for $\beta = 0.5$, and brown for $\beta = 1$.}
	\label{Pic2}
\end{figure*} 

Having $\Delta(t)$, we now determine ground state population of the probe at time $t$. Differentiating Eq.~\eqref{pop} and using Eq.~\eqref{y}, we get
\begin{equation}
    \ddot{\rho}_S^{00} + \Gamma_T \dot{\rho}_S^{00} = -2J^2 \Delta(t).
\end{equation}
Thus, $\rho_S^{00}$ should be obtained by solving this second-order differential equation. The general solution of which can be written as
\begin{equation*}
    \rho_S^{00} = \rho_H + \rho_P,
\end{equation*}
where $\rho_H$ and $\rho_P$ denote the homogeneous and particular solutions, respectively.

We first determine the homogeneous solution $\rho_H(t)$ by setting the source term to zero, which yields
\begin{equation*}
    \rho_H(t) = K_1 + K_2 e^{-\Gamma_T t},
\end{equation*}
where $K_1$ and $K_2$ are constants to be fixed by the initial conditions.

Next, we consider a particular solution of the form
\begin{equation*}
    \rho_P(t) = e^{-\Gamma_T t}\big(a \cos(\Omega t) + b \sin(\Omega t)\big),
\end{equation*}
 Substituting this into the differential equation, we obtain
\begin{equation*}
    a = \frac{2J^2 \Delta_0 \big(\Omega^2 - \Gamma_T^2\big)}{\Omega^2(\Omega^2 + \Gamma_T^2)}, 
    \qquad
    b = \frac{4J^2 \Gamma_T \Omega \Delta_0}{\Omega^2(\Omega^2+ \Gamma_T^2)}.
\end{equation*}

Imposing the initial conditions explicitly at $t=0$, namely $\rho_S^{00}(0) = \rho_i$ (with $\rho_i$ given in Eq.~\eqref{16}) and $\dot{\rho}_S^{00}(0) = 0$, we obtain the final form of the probe state
\begin{equation*}
    \rho_S(\beta,t) = \rho_S^{00}(t)\ketbra{0}{0} + \rho_S^{11}(t)\ketbra{1}{1},
\end{equation*}
where
\begin{equation}
\begin{aligned}
\rho_S^{00}(t) &= \left(1 - p - \frac{b\Omega}{\Gamma_T}\right) 
+ \left(\frac{b\Omega}{\Gamma_T} - a\right)e^{-\Gamma_T t} \\
&\quad + e^{-\Gamma_T t}\big(a \cos(\Omega t) + b \sin(\Omega t)\big),
\label{transstate}
\end{aligned}
\end{equation}
and $\rho_S^{11}(t) = 1 - \rho_S^{00}(t)$.

The functional dependence of $\rho_S(\beta,t)$ on the bath inverse temperature $\beta$ arises through $\Gamma_T$, which is explicitly $\beta$-dependent. One can therefore use $\rho_S(\beta,t)$ to evaluate the quantum Fisher information (QFI) with respect to $\beta$ at a given time $t$.

Finally, it can be verified by direct calculation that in the long-time limit $t \to \infty$, $\rho_S(\beta,t) \to \tau_{\beta}$. Consequently, the steady-state QFI coincides with that of the thermal state at inverse temperature $\beta$.

We now address the following question: In the present setup, the reduced dynamics of the probe is no longer guaranteed to be Markovian due to the auxiliary mediated interaction, which effectively acts as a memory reservoir and allows for information backflow from the auxiliary to the probe. Does the bias toward initially colder probe states, leading to enhanced precision in the transient regime as observed in Markovian dynamics in Sec.~\ref{mar}, still persist? In other words, how is this bias affected by the presence of memory effects?

 To investigate this, we define the ratio $\beta_r := \beta / \beta_i$. For initially colder probes, $0 < \beta_r < 1$, whereas for initially hotter probes, $0 < \beta_r^{-1} < 1$. We consider both regimes and evaluate the QFI at transient times. 

We first fix $\beta$ and $\beta_r$, which in turn determines $\beta_i$. Using $\beta_i$, the initial state $\rho_i$ is obtained. Substituting the ground-state population of $\rho_i$, we then obtain $\rho_S(\beta,t)$ using Eq.~\eqref{transstate}. 

Next, we compute the QFI using Eq.~\eqref{QFM} and evaluate its maximum over time, defined as
\begin{equation}
\mathcal{F}_M = \max_{t} \, \mathcal{F}\big(\rho_S(\beta,t)\big),
\end{equation}
where $\mathcal{F}\big(\rho_S(\beta,t)\big)$ denotes the QFI with respect to $\beta$.

For a probe-auxiliary interaction strength $J = 10$, with $\omega = 1$ and $\gamma = 10^{-4}$, we plot $\mathcal{F}_M$ against $\beta_r$  for both initially hotter and $\beta_r^-1$ for initially colder probes. The left panel of Fig.~\ref{Pic2} shows the behavior of $\mathcal{F}_M$ for initially hotter probes. As evident from the plots, for all considered values of $\beta$, the maximum QFI remains constant and coincides with that of the thermal state corresponding to the same inverse temperature $\beta$. We have verified that this behavior persists over a wide range of $\beta$. This indicates that initially hotter probes do not provide any enhancement in estimation precision, in the transient regime.

In contrast, the right panel of Fig.~\ref{Pic2} corresponds to initially colder probe states. In this case, a clear enhancement of the $\mathcal{F}_M$ is observed at finite times. The maximum QFI is highest for the coldest pobe and decreases monotonically as the initial temperature approaches that of the bath, eventually converging to the thermal-state value. This demonstrates that initializing the probe at a temperature lower than that of the bath is essential for achieving enhanced precision in the transient regime, even in the presence of non-Markovian memory effects. 

Thus, under auxiliary-mediated interactions, the initial bias in the probe temperature persists, in the sense that initially colder probes remain necessary for achieving enhanced precision in the transient regime. A natural question, however, is whether this condition is also sufficient. In other words, does initializing the probe at a temperature lower than that of the bath always guarantee an enhancement in precision?

Insight into this question can be gained from the right panel of Fig.~\ref{Pic2}. As observed, for sufficiently large values of $\beta_r$, the maximum quantum Fisher information (QFI), $\mathcal{F}_M$, approaches very closely the QFI of the corresponding thermal state, $\mathcal{F}_{\beta}$. The difference between these two quantities is extremely small, typically of the order of $10^{-9}$ within our numerical resolution. Although $\mathcal{F}_M$ remains marginally larger than $\mathcal{F}_{\beta}$, this difference decreases progressively as $\beta_r \to 1$ and also diminishes with increasing $\beta$, even for small $\beta_r$.

These observations indicate that, while colder initial probes consistently yield a QFI that is not smaller than that of the thermal state, establishing sufficiency requires resolving exceedingly small differences over a continuous range of parameters. In particular, if one adopts a numerical precision threshold of $10^{-8}$, the observed differences become indistinguishable within numerical accuracy. Under this criterion, one can therefore conclude that colder probes are not sufficient to guarantee a meaningful enhancement in precision within this setup.

%Such distinctions lie at the limit of numerical precision and are therefore not reliably distinguishable within the present analysis. Consequently, while our results strongly support the necessity of colder probes for enhanced precision, the question of sufficiency remains inconclusive within numerical accuracy.

\begin{figure}
    \centering
    \hspace{-1cm}
\includegraphics[scale=0.8]{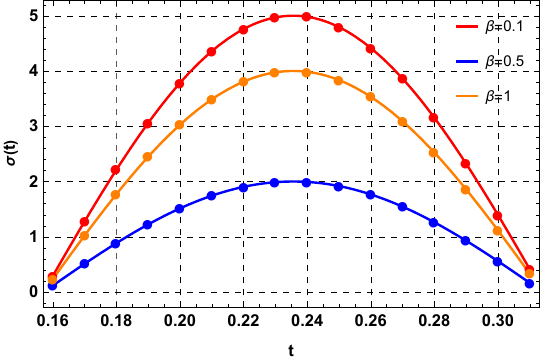}
\caption{\textbf{Evidence of non-Markovianity.} The plot shows the time dependence of $\sigma(t)$, defined in Eq.~\eqref{BLP}, for a representative pair of initial states and for three values of the bath inverse temperature, $\beta = 0.1$, $0.5$, and $1$ (red, blue, and orange, respectively). As seen in the plot, $\sigma(t)$ attains positive values for all considered $\beta$, indicating a breakdown of CP-divisibility and thereby confirming the non-Markovian nature of the dynamics.}
    \label{Pic3}
\end{figure}

Another crucial aspect of our analysis is to establish that the reduced dynamics under consideration is genuinely non-Markovian. At first glance, one might expect non-Markovian behavior since the probe interacts with an auxiliary system, allowing correlations to build up between them. Such correlations can, in principle, lead to a breakdown of CP-divisibility of the reduced dynamics. However, the mere presence of system–environment correlations is only a sufficient condition for non-Markovianity and not a necessary one. For instance, standard Lindblad dynamics is CP-divisible and hence Markovian, even though it may generate transient system–environment correlations.

To rigorously characterize non-Markovianity, we employ the Breuer–Laine–Piilo (BLP) measure~\cite{BLP}, which is based on the behavior of the trace distance between two quantum states. For two initial states $\rho_1(0)$ and $\rho_2(0)$ undergoing the same dynamical map, the BLP measure is defined as
\begin{equation*}
\mathcal{B} = \max_{\rho_1,\rho_2} \int_{\sigma(t) > 0} \sigma(t)\, dt,
\end{equation*}
where
\begin{equation}
\sigma(t) = \frac{d}{dt} D\big(\rho_1(t), \rho_2(t)\big), 
\qquad
D(\rho_1,\rho_2) = \frac{1}{2} \|\rho_1 - \rho_2\|_1.
\label{BLP}
\end{equation}

For CP-divisible (Markovian) dynamics, the trace distance is monotonically non-increasing for all pairs of initial states, implying $\sigma(t) \le 0$ for all $t$. Physically, this reflects a continuous loss of distinguishability due to information flow from the system to the environment. In contrast, a temporary increase in trace distance, i.e., $\sigma(t) > 0$, signals a backflow of information from the environment to the system and thus constitutes a signature of non-Markovianity.

Importantly, although the definition of $\mathcal{B}$ involves a maximization over all pairs of initial states and integration over all time intervals where $\sigma(t) > 0$, the detection of non-Markovianity does not require performing this full optimization. It suffices to identify a single pair of initial states and a time interval for which $\sigma(t) > 0$. This is because CP-divisibility guarantees monotonic contractivity of the trace distance for all pairs of states. Therefore, any violation of this monotonicity for even one pair directly implies that the dynamical map is not CP-divisible and hence non-Markovian in the BLP sense.

In Fig.~\ref{Pic3}, we plot $\sigma(t)$ for a representative pair of initial states $\rho_1(0)$ and $\rho_2(0)$ and for different values of $\beta = 0.1, 0.5, 1$. The function $\sigma(t)$ exhibits oscillatory behavior with multiple intervals where it becomes positive. As a representative example, we highlight a specific time window in which $\sigma(t) > 0$, thereby establishing the presence of information backflow. This demonstrates that, for the chosen parameters $J = 10$, $\gamma = 10^{-4}$, and $\omega = 1$, the reduced dynamics of the probe is non-Markovian.

Thus, our analysis suggests that even in the presence of environmental memory, for certain setups, initially colder probes remain essential for achieving enhanced precision at transient times. However, in the next subsection, we demonstrate that when the dynamics is modeled using a quantum collisional framework, strong environmental memory suppresses this transient precision enhancement, irrespective of the initial probe temperature. This indicates a lifting of the bias between hotter and colder probes, placing them on an equal footing, with neither offering any advantage in precision enhancement.

\section{Breakdown of the Temperature Bias under Non-Markovian Collisional Dynamics}
\label{noncol}
In this section, we show that temperature bias is not a universal feature of transient quantum thermometry. In fact, under strongly non-Markovian dynamics, such a bias can be completely removed. That is, neither initially colder probes nor initially hotter probes provide any advantage over the thermal-state precision.

For this analysis, we consider the non-Markovian collisional model discussed in Sec.~\ref{qc}, and assume the system-auxiliary interaction to be of the Jaynes--Cummings type, given by
\begin{equation}
H_{SA_i}=\Omega\left(\sigma^S_{-}\sigma^{A_i}_{+}+\sigma^S_{+}\sigma^{A_i}_{-}\right),
\end{equation}
where $\sigma^{(\cdot)}_{-}$, with ``$\cdot$'' denoting either $S$ or $A_i$, corresponds to the lowering operator $\ketbra{0}{1}$. As shown in Ref.~\cite{ColmemFt}, for this type of interaction, the map $\mathcal{E}(\rho)$ reduces to the amplitude damping channel when the ancilla is initially prepared in the ground state $\ketbra{0}{0}$. 

However, here we consider the ancilla to be initially in a thermal state of the form
\begin{equation}
\sigma_{A_i}=\frac{e^{-\beta H_{A_i}}}{Z},
\end{equation}
with Hamiltonian $H_{A_i}=\omega\sigma_z$ and partition function
\begin{equation}
Z=\Tr\left[e^{-\beta H_{A_i}}\right].
\end{equation}

The corresponding dynamical map is then given by
\begin{equation*}
\begin{split}
\mathcal{E}_{t}^{\beta}(\rho)
&=\sum_{k}\frac{e^{-\beta E_k}}{Z}
\left(\sum_{j}\bra{j}e^{-i H_{SA_i} t}\ket{k}\,\rho\,\bra{k}e^{i H_{SA_i} t}\ket{j}\right) \\
&=\sum_{l=0}^{3}K_l \rho K_l^\dagger,
\end{split}
\end{equation*}
where $j,k=0,1$ label the ancilla basis states, $E_k$ are the eigenvalues $H_{A_i}$, corresponding to the eigenstate $\ket{k}$ and $K_l$ are the Kraus operators.

Substituting the explicit form of $H_{SA_i}$, the Kraus operators are obtained as
\begin{equation*}
\begin{split}
K_0&=\sqrt{\frac{e^{-\beta E_0}}{Z}}
\left(\ketbra{0}{0}+\cos(\Omega t)\ketbra{1}{1}\right),\\
K_1&=\sqrt{\frac{e^{-\beta E_0}}{Z}}
\sin(\Omega t)\ketbra{0}{1},\\
K_2&=\sqrt{\frac{e^{-\beta E_1}}{Z}}
\sin(\Omega t)\ketbra{1}{0},\\
K_3&=\sqrt{\frac{e^{-\beta E_1}}{Z}}
\left(\cos(\Omega t)\ketbra{0}{0}+\ketbra{1}{1}\right).
\end{split}
\end{equation*}

This has exactly the form of a generalized amplitude damping (GAD) channel with damping parameter $\sin^2(\Omega t)$. Hence, for perfect swapping ($P=1$), the state of the probe at time $t$ is given by
\begin{equation}
\rho(\beta,t)=\mathcal{E}_{t}^{\beta}(\rho_i).
\end{equation}

We consider the initial probe state $\rho_i$, as given in Eq.~\eqref{16}. Substituting $\mathcal{E}_{t}^{\beta}(\rho_i)$, we obtain
\begin{equation}
\rho(\beta,t)=p_0(\beta,t)\ketbra{0}{0}+p_1(\beta,t)\ketbra{1}{1},
\end{equation}
where
\begin{eqnarray}
p_0(\beta,t)&=&(1-p)+(p-p^e)\sin^2(\Omega t),\\
p_1(\beta,t)&=&1-p_0(\beta,t).
\end{eqnarray}

Where $p^{e}$, is given by Eq.~\eqref{therm}. The QFI at time $t$ is then given by
\begin{equation}
\mathcal{F}(\beta,t)=
\frac{\left(\frac{\partial p_0(\beta,t)}{\partial \beta}\right)^2}
{p_0(\beta,t)\,p_1(\beta,t)}.
\end{equation}

Note that $p_0(\beta,t)$ is an oscillatory function with period $\Omega t=\pi$. To determine the maxima of the QFI, we solve $\frac{\partial \mathcal{F}(\beta,t)}{\partial t}=0$, which yields three conditions:
\begin{itemize}
\item $\sin^2(\Omega t)=0$,
\item $\sin(2\Omega t)=0$,
\item $\sin^2(\Omega t)=\frac{2(1-p)p}{(1-2p)(p-p^e)}$.
\end{itemize}

For the first and third conditions, $p_0(\beta,t)$ becomes independent of $\beta$, and therefore the corresponding QFI vanishes. In contrast, under the second condition, the QFI is exactly equal to that of the thermal state, since the probe state becomes
\begin{equation}
\rho\!\left(\frac{\pi}{2\Omega}\right)
=(1-p^e)\ketbra{0}{0}+p^e\ketbra{1}{1}.
\end{equation}

At the boundary points $t=n\pi/\Omega$, with $n=1,2,3,\dots$, the QFI also vanishes. Therefore, the maximum QFI attainable under the perfect-swap condition in the quantum collisional model is exactly equal to the thermal-state QFI.

This analysis shows that, although in the Markovian limit (corresponding to the no-swap condition in the quantum collisional model) there is a transient precision enhancement when the probe is initially colder than the bath, such an enhancement is completely absent in the strongly non-Markovian regime (perfect swap). In this regime, both colder and hotter probes are placed on equal footing, with neither capable of achieving enhanced precision.

In the next section, we conclude our analysis.
\section{Conclusion and Discussion}
In this work, we investigated how the initial preparation of a quantum probe influences the achievable precision in transient quantum thermometry. Focusing on qubit probes undergoing Markovian dynamics while interacting with a bosonic thermal bath, we identified a clear and universal operational criterion governing transient precision enhancement. Specifically, we showed that enhanced estimation precision at finite times can be achieved if and only if the probe is initially prepared in a state that is colder than the thermal state corresponding to the bath temperature to be estimated. Remarkably, this condition is independent of the specific choice of the probe Hamiltonian energy gap, highlighting its universal character.

We further examined how the nature of this bias is modified in the presence of environmental memory, thereby incorporating non-Markovian effects into the dynamics. In particular, we considered two distinct scenarios. In the first case, non-Markovianity was induced via auxiliary-mediated interactions. In the second case, we analyzed a quantum collisional model under the perfect swap condition.

We found that in the first scenario, achieving an enhancement in precision over the steady-state value still requires the probe to be initialized at a temperature lower than that of the bath. In contrast, in the second scenario, there is a complete absence of transient precision enhancement, irrespective of the choice of the initial probe state. In this case, both hotter and colder initial probes are placed on equal footing, as neither is capable of providing an advantage in transient precision. This indicates a complete disappearance of the bias observed in the Markovian regime.

Thus our results provide a clear and operationally meaningful framework for quantum thermometry in the transient regime. In the Markovian setting, we identify a simple and robust principle: preparing the probe in a colder state guarantees enhanced precision at finite times, independent of microscopic details such as the energy gap. This offers a practical guideline for designing efficient quantum thermometers with assured performance advantages. Furthermore, our analysis of non-Markovian dynamics showed that environmental memory can significantly alter, and in some cases completely suppress, this advantage. Understanding how memory effects influence the bias toward colder probes is crucial from a practical standpoint, as it determines the regimes in which transient thermometric enhancement remains viable.

%We further demonstrated that the apparent temperature bias is not a fundamental limitation of quantum thermometry, but rather a consequence of Markovian dynamics. By introducing non Markovianity through coupling the probe to an auxiliary system that mediates its interaction with the bath, we showed that the restriction imposed by the initial probe temperature can be lifted. In particular, in the non Markovian regime, the dependence of the maximum achievable precision on the probe’s initial temperature is mitigated, with probes prepared both hotter and colder than the bath attaining the same optimal precision at finite times. Moreover, this transient enhancement surpasses the steady state precision reached asymptotically.

%Our results thus establish a sharp operational distinction between Markovian and non-Markovian quantum thermometry. While Markovian dynamics impose a strict condition on the initial probe temperature for achieving transient enhancement, non-Markovian dynamics relax this constraint and allow enhanced performance regardless of the initial probe bias. These insights can guide the design of optimal quantum thermometers that ensure enhanced precision in the presence of Markovian dynamics.
\label{con}
\bibliography{therm}
\end{document}